\documentclass{aa}

\usepackage{txfonts}
\usepackage{graphicx}
\usepackage{dcolumn}
\usepackage{bm}
\usepackage{xspace}
\usepackage{xcolor}
\usepackage{array}
\usepackage{hyperref}
\usepackage[mathlines]{lineno}
\usepackage{physics}
\usepackage{booktabs}
\usepackage[normalem]{ulem}

\newcommand{\AREPO}{\textsc{arepo}\xspace}
\newcommand{\msun}{\,$M_\odot$\xspace}
\newcommand{\msol}{\,$M_\odot$\xspace}

\newcommand{\rsol}{\,$R_\odot$\xspace}
\newcommand{\E}[1]{\,{\times}\,10^{#1}\xspace}

\makeatletter
\renewcommand*\aa@pageof{, page \thepage{} of \pageref*{LastPage}}
\makeatother

\begin{document}

\title{Faint calcium-rich transient from a double-detonation of a 0.6\msol carbon-oxygen white dwarf star}
\titlerunning{Double detonation of a 0.6\msol CO WD}
\author{
    Javier Mor\'an-Fraile\inst{1}
    \and
    Alexander Holas\inst{1,2}
    \and
    Friedrich K. Röpke\inst{1,2}
    \and
    Rüdiger Pakmor\inst{3}
    \and
    Fabian R. N. Schneider\inst{1,4}
    }

\institute{
    Heidelberger Institut für Theoretische Studien (HITS),
              Schloss-Wolfsbrunnenweg 35, 69118 Heidelberg, Germany
    \and
    Zentrum f\"ur Astronomie der Universit\"at Heidelberg,
    Institut f\"ur Theoretische Astrophysik, 
    Philosophenweg 12,
    69120 Heidelberg, Germany
    \and
    Max-Planck-Institut f\"ur Astrophysik,
    Karl-Schwarzschild-Str. 1, 85748 Garching,
    Germany
    \and
    Zentrum f{\"u}r Astronomie der Universit{\"a}t Heidelberg,
    Astronomisches Rechen-Institut,
    M{\"o}nchhofstr.\ 12-14, 
    69120 Heidelberg, Germany
    }

\date{\today}

\abstract{
We have computed a three-dimensional hydrodynamic simulation of the merger between a massive ($0.4$\msol) helium white dwarf (He WD) and a low-mass ($0.6$\msol) carbon-oxygen white dwarf (CO WD). Despite the low mass of the primary, the merger triggers a thermonuclear explosion as a result of a double detonation, producing a faint transient and leaving no remnant behind. This type of event could also take place during common-envelope mergers whenever the companion is a CO WD and the core of the giant star has a sufficiently large He mass. The spectra show strong Ca lines throughout the first few weeks after the explosion. The explosion only yields $<0.01$\msol of $^{56}$Ni, resulting in a low-luminosity SN Ia-like lightcurve that resembles the Ca-rich transients within this broad class of objects, with a peak magnitude of $M_\mathrm{bol} \approx -15.7$\,mag and a rather slow decline rate of $\Delta m_{15}^\mathrm{bol}\approx 1.5$\,mag. Both, its lightcurve-shape and spectral appearance, resemble the appearance of Ca-rich transients, suggesting such mergers as a possible progenitor scenario for this class of events.}

\maketitle

\section{\label{sec:intro}Introduction}
Binary white dwarf (WD) systems are the most common type of double compact objects in our galaxy \citep{nelemans2001a}. 
Consequently, their mergers are a common occurrence.
These mergers can take place  because gravitational wave emission (GW), common-envelope events, three-body interaction,  and other processes can shrink the orbit of the binaries down to the point of Roche-lobe overflow (RLOF), giving rise to a variety of astronomical transients depending on the mass and composition of the involved WDs.
These mergers have been primarily studied with the goal of finding the progenitors of Type Ia supernovae \citep[SNe~Ia,][]{liu2023}, which are thought to be produced by thermonuclear explosions in carbon-oxygen (CO) WDs.
To enhance the chance for the ignition of a thermonuclear explosion and to ensure the production of the ${\sim}\, 0.5$\msol of $^{56}\mathrm{Ni}$ required to power the lightcurve of a normal SN~Ia, most of the previous research has focused on mergers involving CO WDs with masses above $0.9$\msol \citep{sim2010a, shen2021a}.
The exact mechanism that initiates the explosion is debated, but most of the proposed scenarios involve a so-called ``double-detonation''.
In this scenario, a helium (He) shell surrounding the CO WD is ignited initially.
The detonation of the He shell converges near the core of the CO WD. In the convergence point, the density and temperature are greatly increased and a detonation wave is launched that propagates outwards incinerating the CO WD material.
But higher-mass CO WDs are less abundant in nature than their lower-mass counterparts both among single stars \citep{kepler2007a} as well as in binaries \citep{nelemans2001a, toonen2012a}.
If low-mass CO WDs were to produce some type of detectable transient, they could potentially be much more frequent than SNe Ia due to their ubiquity.
At masses ${\sim}\,0.6$\msol, the central density of CO WDs is one order of magnitude below that of their $0.9$\msol counterparts, making the conditions for carbon ignition \citep{seitenzahl2009b} difficult to attain, even under the double-detonation scenario.
For these low-mass CO WDs to explode, a new mechanism for their detonation must be found or they must be enveloped in a very massive He shell that generates a more potent compression than in the classic double-detonation scenario, compressing the CO WD core enough to reach the conditions of carbon ignition.

Only material burned at densities over $\rho\,{>}\,10^7\textrm{g}\,\textrm{cm}^{-3}$ is able to produce $^{56}\textrm{Ni}$ \citep[see, e.g.,][]{sim2010a}, and therefore any potential explosion of a low-mass CO WD would result in a very low $^{56}\textrm{Ni}$ yield. As a direct consequence of the low $^{56}\textrm{Ni}$ yield, these explosions would produce transients much fainter than SN Ia.

Here we present a three-dimensional magnetohydrodynamic (3D MHD) simulation of the merger between a $0.4$\msol He WD and a $0.6$\msol CO WD  exploding in a double detonation scenario with no bound remnant. The outcome is a faint transient that shares features with Ca-rich transients \citep{perets2010a, waldman2011a, kasliwal2012a, taubenberger2017a, shen2019a}. Throughout this paper we use the term ``Ca-rich'' but we note that it is debated whether a substantial amount of calcium is required to excite the Ca line in the late spectra\citep{shen2019a,jacobson2020}.

The thermonuclear explosion takes place minutes after the disruption of the He WD as the disrupted material is accreted by the CO WD.

\section{Methods\label{sec:methods}}
\subsection{Hydrodynamic explosion simulation}
The hydrodynamic simulation is carried out with the \AREPO code \citep{springel2010a,pakmor2011d,pakmor2016a,weinberger2020a}, which is a 3D MHD code that uses a second-order finite volume approach on an unstructured moving Voronoi mesh.
In our simulation, we employ similar methods as in \citet{pakmor2021a}.
We use explicit refinement and de-refinement for cells that are more than a factor of two away from the target mass of the cells ($m_\textrm{target} = 5\times10^{-7}~\textrm{M}_\odot$), which results in $\sim 2\E{6}$ \AREPO cells.
Self-gravity is treated as Newtonian and computed using a oct-tree based algorithm. We use the Helmholtz equation of state \citep{timmes2000a} including Coulomb corrections.
We also include a 55 isotope nuclear reaction network coupled to hydrodynamics \citep{pakmor2012b,pakmor2021a,gronow2021a} that includes the following isotopes: n, p, $^4$He, $^{11}$B, $^{12-13}$C, $^{13-15}$N, $^{15-17}$O, $^{18}$F, $^{19-22}$Ne, $^{22-23}$Na, $^{23-26}$Mg, $^{25-27}$Al, $^{28-30}$Si, $^{29-31}$P, $^{31-33}$S, $^{33-35}$Cl, $^{36-39}$Ar, $^{39}$K, $^{40}$Ca, $^{43}$Sc, $^{44}$Ti, $^{47}$V, $^{48}$Cr, $^{51}$Mn, $^{52,56}$Fe, $^{55}$Co, and $^{56,58-59}$Ni.
We use the JINA reaction rates \citep{cyburt2010a}. Nuclear reactions are computed for all cells with $T > 2\times10^7~\textrm{K}$, but we limit burning at shock front as described by \citet{pakmor2021a}.

\subsection{Setup}
Our initial binary system consists of a $0.6$\msol WD with an equal-by-mass mixture of carbon and oxygen and an initial radius of $0.0121$\rsol\footnote{Defined as the radius in which $99.9\%$ of the stellar mass is contained}, and a $0.4$\msol WD composed entirely of He with a radius of $0.0156$\rsol. Our simulation is set up similar to that of \citet{pakmor2021a}.
The stellar models are constructed in hydrostatic equilibrium assuming a uniform temperature of $\textrm{T}=10^6\,\mathrm{K}$.
We map the density profiles to a 3D mesh in \AREPO and assign the cells the appropriate internal energy for the given temperature using the Helmholtz equation of state. The stars are first relaxed separately to ensure hydrostatic equilibrium and remove noise introduced by the mapping process.
The relaxation procedure is the same for both stars: they are placed in a cubic box of $2\times10^{10}\,\mathrm{cm}$ sides, with a uniform background grid density of $\rho = 10^{-5}\,\mathrm{g\,cm}^{-3}$ and a thermal energy of $10^{14}\,\textrm{erg}\,/\textrm{g}$  ($\textrm{T}\,{\sim}\,5\E{5}\,\textrm{K}$).
The duration of the relaxation is chosen to be ten times the dynamical timescale of each star, i.e.\ $20\,\textrm{s}$ for the $0.6$\msol CO WD and $37\,\textrm{s}$ for the $0.4$\msol He WD.
During the first half of the relaxation, we damp the spurious velocities  introduced by discretization errors.
During the second half of the relaxation, the stars are left to evolve freely without the damping force, and we check that the density profiles remain unchanged.

After the relaxation, we place both stars in a box with a size of $7.1\E{11}\,\mathrm{cm}$ and uniform background density of $\rho = 10^{-5}\,\mathrm{g\,cm}^{-3}$. To avoid large tidal forces when putting the WDs together in the box, we chose the initial distance between the WDs to be $9.4\E{9}\,\textrm{cm}$,   which is equal to three times the separation at the point of RLOF. 
We then artificially shrink the orbit on a timescale larger than the dynamical timescales of the WDs so that potential tidal effects can still take place.
This orbital shrinking follows the scheme described in \citet{pakmor2021a}, and it is stopped at the point of RLOF, when a steady mass transfer sets in.
This happens at an orbital separation of $3.14\E{9}\,\textrm{cm}$, with a total angular momentum of $L=3.2\E{50}\,\textrm{g\,cm}^2\,\textrm{s}^{-1}$. From this point on, we let the system evolve freely.
We run the simulation with the nuclear network disabled during the artificial inspiral in order to reduce the computational cost, as no significant fraction of mass is expected to undergo any nuclear reaction during this stage.
We enable the nuclear network once the artificial orbital shrinking is terminated.

\subsection{Prediction of observables}
For a precise computation of the optical spectrum emitted by the event, we postprocess the simplified nucleosynthesis of our \AREPO run by following the thermodynamic trajectories of $10^6$ tracer particles during the explosion and using them as an input for a 384 isotope nuclear reaction network \citep{seitenzahl2010a,pakmor2012a,seitenzahl2017a} to obtain more accurate nucleosynthetic yields in the ejecta and assuming solar metallicity \citep{asplund2009a}.
The explosion is simulated in \AREPO until the ejecta reach homologous expansion.
At this point, the ejecta structure as determined in the nucleosynthetic postprocessing step is mapped to the three-dimensional Monte Carlo radiative transfer code ARTIS \citep{sim2007b, kromer2009a, bulla2015a, shingles2022a} following the methods described by \citet{gronow2020a} to derive synthetic observables.
The postprocessed yields of the ejecta structure are mapped to a $50^3$ Cartesian grid on which the Monte Carlo radiation transfer is computed. We track $10^8$ energy packets as they propagate through the ejecta for $100$ logarithmically spaced timesteps between $2$ and $60$ days after the explosion.

\section{Results\label{sec:results}}
\subsection{Merger and double detonation}
\begin{figure*}
    \centering
    \includegraphics[width=\textwidth]{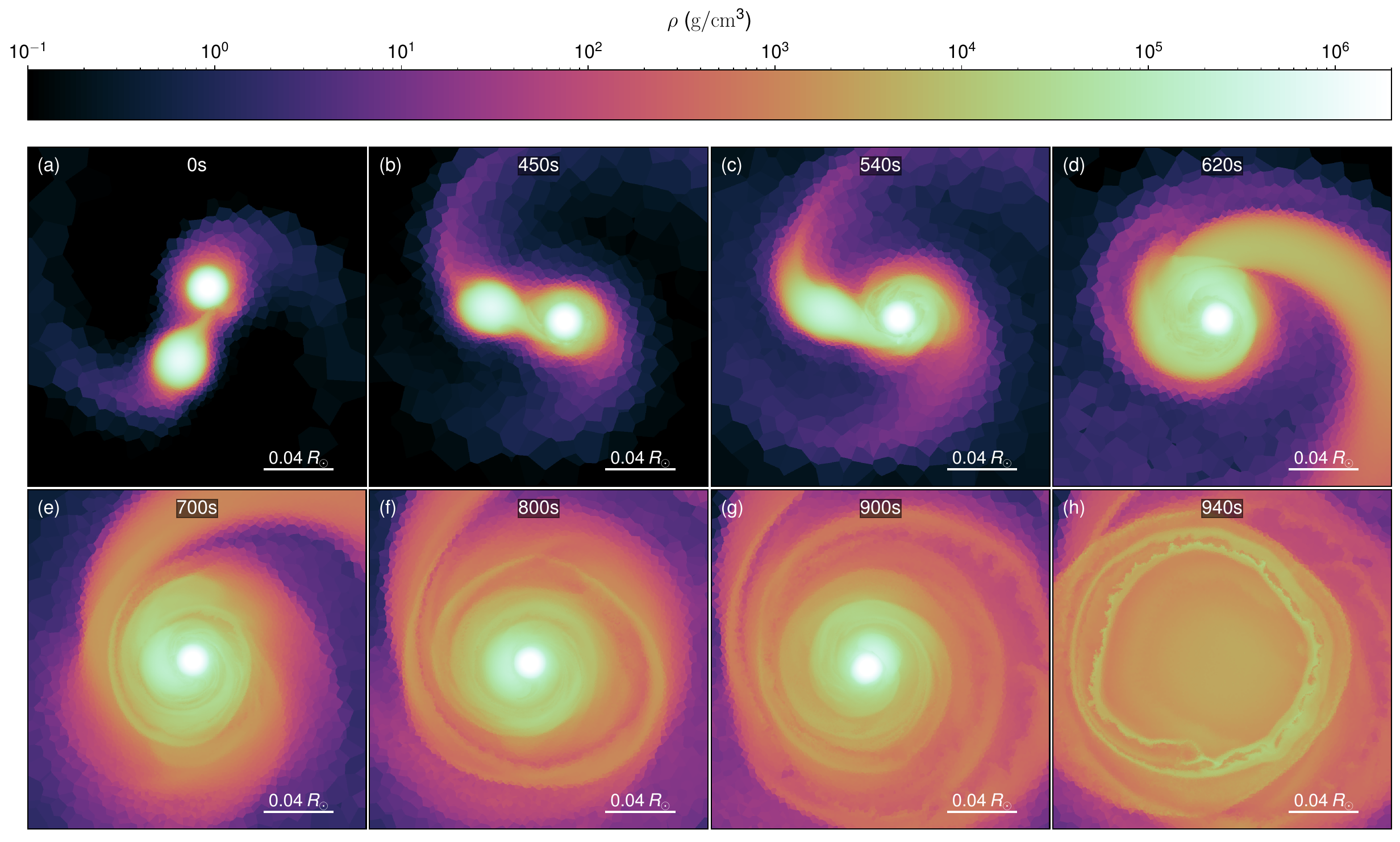}
    \caption{
    Hydrodynamic evolution of the WD-WD merger. The panels show slices of the orbital plane color coded with the density from the beginning of the interaction until the moment that the double detonation takes place.
}
    \label{fig:merger}
\end{figure*}

The evolution of the system is visualized in Fig.~ \ref{fig:merger} and in the complimentary movies\footnote{available on Zenodo (\url{https://doi.org/10.5281/zenodo.8268166}, showing the temperature, magnetic fields, temperature and kinetic pressure)}. Both stars are initialized with a dipolar magenetic field with a value of $10\,\textrm{G}$ at the equator; nevertheless, magnetic fields play a minor role in this simulation.
When reaching the point of RLOF, He is transferred from the larger, less massive He WD onto the surface of the CO WD.
Being composed of fully degenerate matter, the He  WD expands as it loses mass, increasing the mass-transfer rate.
This shrinks the orbit in a runaway process that ends when the He  WD is tidally disrupted by the CO WD after $550\,\textrm{s}$.
The disrupted stellar material initially forms a spiral arm, efficiently transferring  angular momentum outwards  (Fig.~\ref{fig:merger}d-g). This allows a large fraction of the disrupted gas to spiral inwards, closer to the CO WD, where it begins to form a non-homogeneous He ``shell'' that completely engulfs the CO WD.
This process is complex and we refer the reader to the complimentary movies for a better understanding.
Due to a tidal interaction, part of the infalling material intersects itself and gets ``trapped'', forming an overdense and cold fluid parcel. While the rest of the accreting material circularizes and heats up due to shear and turbulent motions, this overdense structure remains colder with respect to the surrounding gas that forms the He ``shell''. As the fluid parcel draws closer to the surface of the CO WD, it gets compressed and increases its density up to $3\E{5}\,\textrm{g\,cm}^{-3}$.

\begin{figure*}
    \centering
    \includegraphics[width=\textwidth]{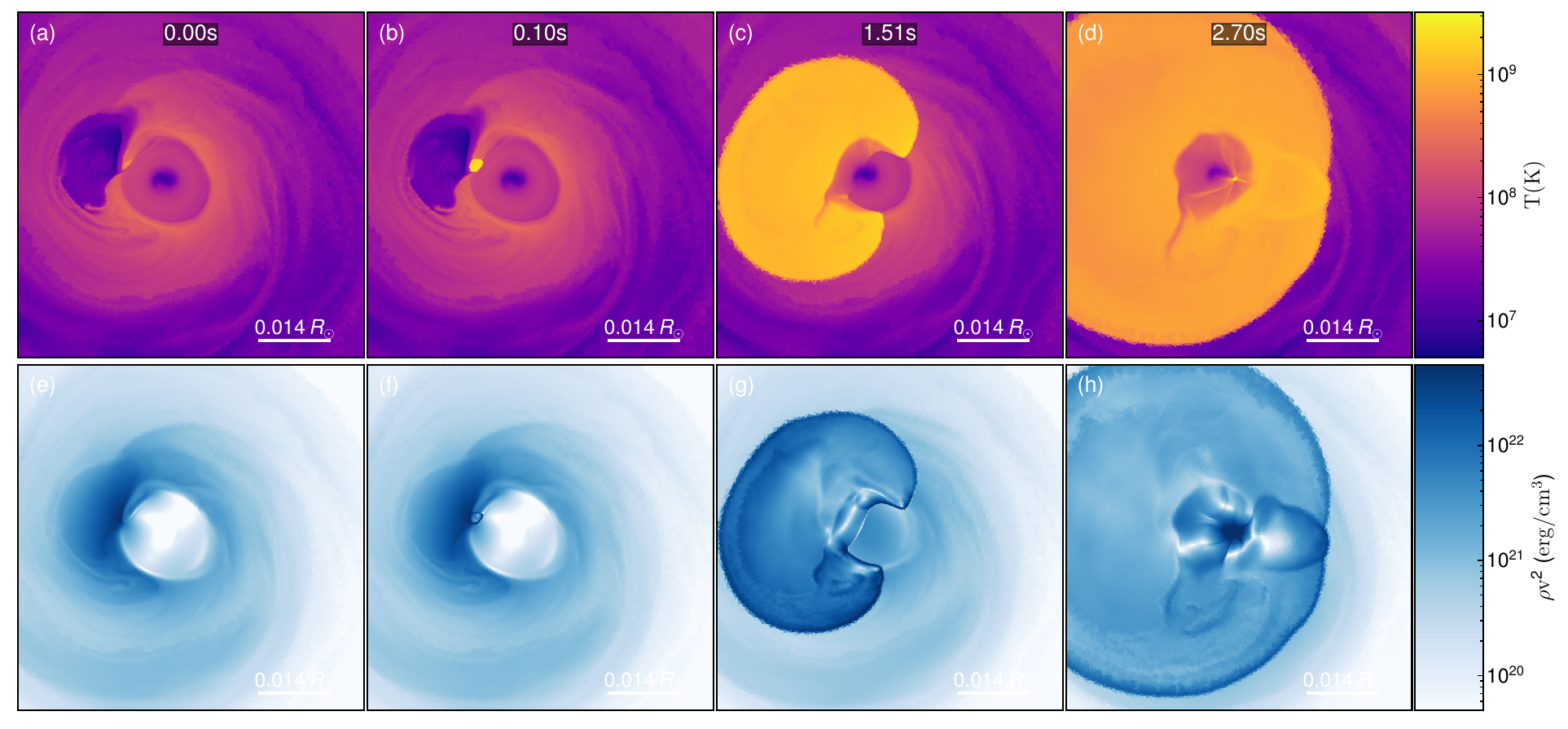}
    \caption{
    Columns show snapshots immediately before the first detonation (t=$0\,\textrm{s}$), first snapshot after the He ignition (t=0.1s) and the propagation of the shock produced by this detonation through the CO WD (t=$1.51\,\textrm{s}$) converging at $2.7\E{8}\,\textrm{cm}$ from the center of the CO WD (t=$2.7\,\textrm{s}$).
    \emph{Top}: Temperature in $\textrm{K}$.
    \emph{Bottom:} Kinetic pressure in $\textrm{erg}/\textrm{cm}^{3}$.
    \label{fig:explosion}
    }
\end{figure*}

When this overdense material reaches the surface of the CO WD, it shears against it (the overdense material is best identified in Fig.~\ref{fig:explosion}a as the colder region), heating up and compressing the contact region. The shear dredges up a small fraction of C and O from the primary that is mixed with He in the ``shell''.
In this shearing layer, the density reaches values of up to $3\E{5}\,\textrm{g\,cm}^{-3}$ and temperatures of $7\E{8}\,\textrm{K}$.

A He detonation takes place $381.5$s after the disruption of the He WD. The detonation is shown in Fig.\ref{fig:explosion}.
The explosion originates from the region between the cold and overdense He structure and the CO WD (Fig.\ref{fig:explosion}b \& Fig.\ref{fig:explosion}f.
The shock propagates through the CO WD heating and compressing the material while the detonation wave propagates around the CO WD (Fig.~\ref{fig:explosion}c and Fig.~\ref{fig:explosion}g) releasing $3\E{50}\,\textrm{erg}$ of nuclear energy (Table~\ref{table:yields}).
The shocked CO does not reach temperatures or densities high enough to trigger burning inside the CO WD as the shock passes through it.
The shock wave converges at a distance of $2.7\E{8}\,\textrm{cm}$ from the center of the CO WD (Fig.~\ref{fig:explosion}d \& Fig.\ref{fig:explosion}h) 2.7s after the first detonation, compressing the material to densities of $\rho = 1.2\E{7}\,\textrm{g\,cm}^3$ and temperatures up to $\textrm{T}=4.5\E{9}\,\textrm{K}$, igniting a carbon detonation that propagates outward incinerating the CO material and releasing $6.8\E{50}\,\textrm{erg}$ (Table~\ref{table:yields}). The CO WD is completely disrupted, with the ejecta having a partial axial symmetry due to the interaction with the disrupted stellar material located mainly on the orbital plane.

 \begin{table}
  \caption[]{Final abundances and energy released by the He and CO detonations, and in the entire event.}
     \label{table:yields}
        \begin{tabular}{p{0.25\linewidth}p{0.17\linewidth}p{0.17\linewidth}p{0.17\linewidth}} \toprule
           & He \newline detonation & CO \newline detonation & Total\\
          \midrule
          Energy released \newline $[10^{50}\,\textrm{erg}]$ & $3.0$ & $3.8$ & $6.8$\\
          \midrule
          Yield $[10^{-3}\,M_\odot]$ & & & \\
          $^4$He & $320$ & $ 0.0 $ & $320$ \\
          $^{12}$C & $4.6$ & $61.6$ & $66.2$\\
          $^{16}$O & $1.1$ & $ 208 $ & $209$ \\
          $^{20}$Ne & $0.4$ & $18.0$ & $18.4$\\
          $^{24}$Mg & $0.7$ & $27.2$ & $27.9$ \\
          $^{28}$Si & $6.7$ & $168$ & $175$\\
          $^{32}$S & $16.7$ & $71.8$ & $88.5$\\
          $^{36}$Ar & $9.3$ & $9.4$ & $18.6$\\
          $^{40}$Ca & $18.8$ & $5.6$ & $24.4$\\
          $^{44}$Ti & $6.3$ & $0.0$ & $6.3$\\
          $^{48}$Cr & $6.0$ & $0.0$ & $6.0$\\
          $^{56}$Ni & $3.3$ & $4.7$ & $8.0$\\
         [0.5ex]
         \bottomrule
        \end{tabular}
\end{table}

\subsection{Ejecta and radiative transfer}
The final yields and explosion energies of the merger are listed in Table \ref{table:yields}. Out of the total ejecta mass of about 1\,$\textrm{M}_\odot$, a substantial amount  (about $33$\%) is composed of He that remain unburnt.
O contributes around $20$\% to the total ejecta mass, whereas the iron group elements (IGEs) only constitute less than $5$\%. The remainder consists of intermediate mass elements (IMEs), with Ca making up around $2.5$\% of the final composition.
The ${\sim}\,0.08$\msol of He that is ignited burns mainly into $^{40}$Ca and $^{32}$S, and this burning is the origin of all the $^{44}$Ti and $^{48}$Cr produced in the simulation.
A small fraction of the burnt He produces $^{12}$C and around $40\%$ of the total $^{56}$Ni yield of the event. After the CO detonation a ${\sim}\,30\%$ of the initial $^{12}$C remains unburned, while the rest of was burnt to $^{16}$O, the majority of the produced IMEs, and $60\%$ of the total $^{56}$Ni yield (Table~\ref{table:yields}).

What stands out in the ejecta composition as listed in Table~\ref{table:yields} is the rather low $^{56}$Ni abundance compared with a substantial amount of $^{40}$Ca. This composition, combined with the low luminosity implied by the low $^{56}$Ni yield, suggests that our model may resemble a Ca-rich transient (see e.g. \cite{taubenberger2017a}). Furthermore, the $^{56}$Ni yield also suggests that this is a sub-luminous event compared with a normal SNe~Ia.

\begin{figure*}
    \centering
    \includegraphics[width=\textwidth]{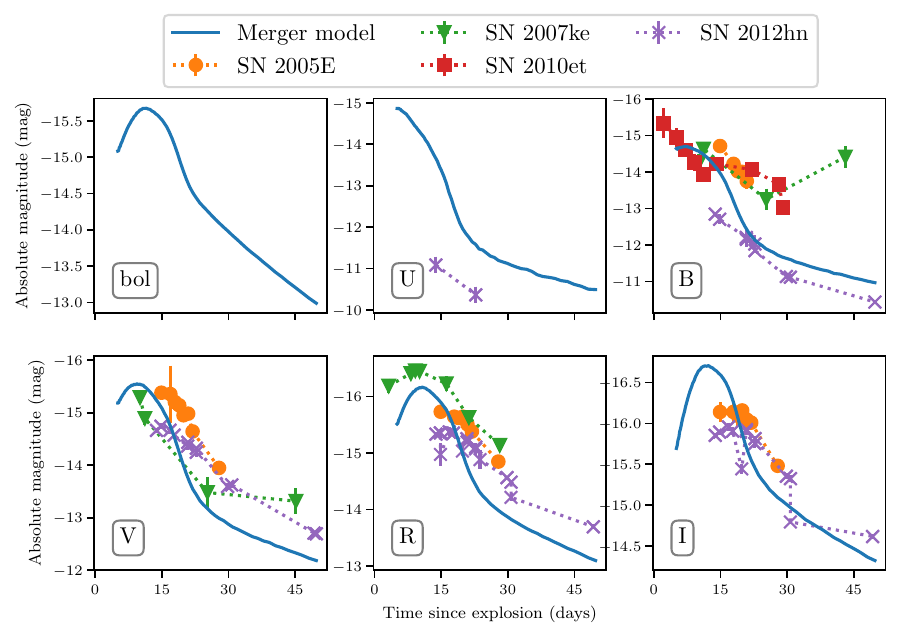}
    \caption{
    Bolometric and broadband light curves of the merger model. Here, several Ca-rich transient candidates are plotted as well for comparison: SN 2005E \cite{perets2010a}, SN 2007ke, SN 2010et \cite{kasliwal2012a} and SN 2012hn \cite{valenti2014a}. All data are taken from the Open Supernova Catalog \cite{guillochon2017a}.
    \label{fig:lightcurves}
    }
\end{figure*}

This is confirmed in our radiative transfer simulations, which predict low-luminosity lightcurves powered by the radioactive decay of $^{56}$Ni (Fig.~\ref{fig:lightcurves}) that resemble the Ca-rich transients within the broader class of SN Ia-like objects. The angle-averaged lightcurves show a bolometric peak magnitude of $M_\mathrm{bol}\,{\approx}\,-15.7$\,mag and a rather slow decline rate of $\Delta m_{15}^\mathrm{bol}\approx 1.5$\,mag.
Furthermore, we observe a substantial viewing angle dispersion in peak magnitudes with values ranging from $M_\mathrm{bol} \approx -15.4$\,mag to $M_\mathrm{bol} \approx -16.0$\,mag and decline rates ranging from $\Delta m_{15}^\mathrm{bol}\approx 1.4$\,mag to $\Delta m_{15}^\mathrm{bol}\approx 1.7$\,mag.
This large scatter can be attributed to the pronounced asymmetry in the ejecta structure as a result of the double detonation explosion mechanism.
A detailed analysis of the viewing angle dependence, however, is beyond the scope of this study and will be discussed in future work.

\begin{figure*}
    \centering
    \includegraphics[width=\textwidth]{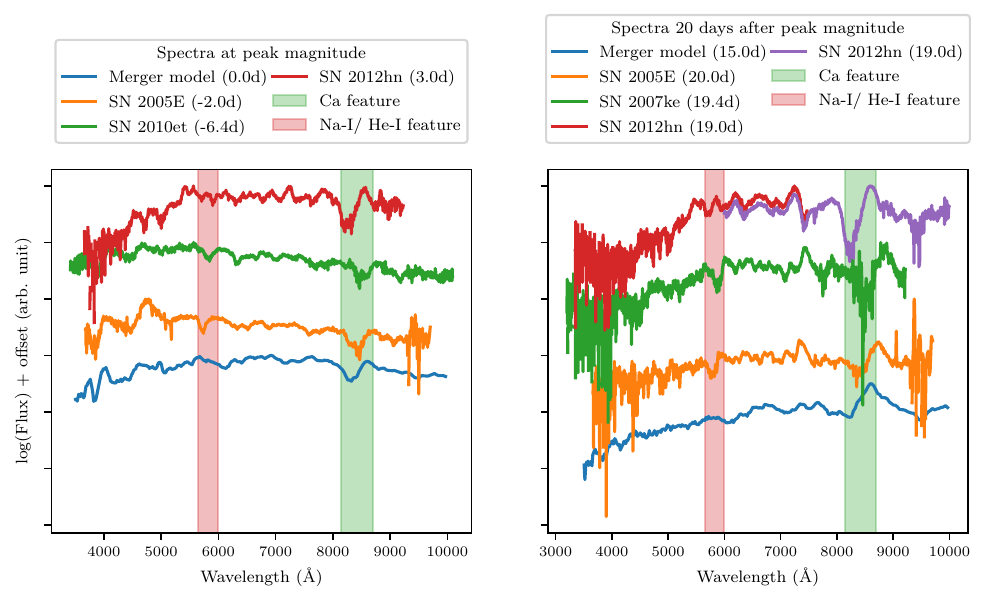}
    \caption{
    Spectral comparison between our merger model and various Ca-rich transient candidates \citep{perets2010a,kasliwal2012a,valenti2014a,shivvers2019a}. Here the prominent calcium feature can be seen (green marking) as well as the poorly reproduced Na-I/ He-I feature (red marking). It should be noted that these markings do not indicate the precise location of the respective features, but rather serve as a guide for the eye. All data has been taken from WISeREP \citep{yaron2012a}. \emph{Left:} Spectra around peak magnitude. \emph{Right:} Spectra around $20$ days after peak magnitude.
    \label{fig:spectra}
    }
\end{figure*}

We compare the resulting broad-band and bolometric light curves to several typical Ca-rich transient candidates in Fig.~\ref{fig:lightcurves}, where find a rather good qualitative agreement, particularly in the \textit{V}- and \textit{R}-band peak magnitudes and decline rates.
However, the spectra obtained from our radiative transfer simulation only somewhat resemble the observed spectra of Ca-rich transients, as is illustrated in Fig.~\ref{fig:spectra}.
While we reasonably reproduce the calcium feature, we consistently lack the feature visible at around $5900$\,\AA, marked in red in Fig.~\ref{fig:spectra}, interpreted as either a He-I or Na-I line.

\section{Discussion and conclusions\label{sec:discussion}}
The cells in the region of He ignition have temperatures of $T\,{\sim}\,7\E{8}\,\textrm{K}$ and densities of $\rho\,{\sim}\,3\E{5}\,\textrm{g\,cm}^{-3}$ with a typical size of  ${\sim}\, 90\,\textrm{km}$. We also find a small fraction of C (about $15\%$) mixed into the composition of the detonating material, which is known to facilitate the He ignition. These conditions are consistent with simulations of resolved He detonations \citep{shen2014b} where the ignition points have length scales of hundreds of kilometers. We see the ignition of He in several cells and consider it marginally resolved in our simulation.
This is not the case for the triggering of the CO detonation, which takes place on centimeter scales \citep{seitenzahl2009b}, i.e.
 far below the typical resolution of ${\sim}\,15\,\textrm{km}$ in the CO detonation ignition region in our simulation. Still, at the point of shock convergence, we measure densities of $\rho = 1.2\E{7}\,\textrm{g\,cm}^3$ and temperatures of up to $\textrm{T}=4.5\E{9}\,\textrm{K}$, which are consistent with those found in simulations of resolved CO detonations \citep{seitenzahl2009b}. 
Therefore we argue that the detonations we observe in our simulation are physically plausible.
One caveat is the simplistic initial composition used for the explosion simulation. More realistic WD models should include $^{22}$Ne in the CO material and $^{14}$N in the He to represent the metallicity of the original star. It has been shown by that these pollutants facilitate the He ignition \citet{shen2014b} and they alter the composition of the burning products.

The mass distribution of CO WDs is skewed towards lower masses, and, due to their low central densities, low-mass WDs  were previously believed to be very unlikely to produce a thermonuclear explosion. We find that the double detonation mechanism resulting from mergers with massive He WDs can lead to an explosion of such low-mass CO WDs. The He WD companions in the merger result from the cores of RG stars whose envelope is removed in binary interaction. Stars born with masses up to ${\sim}\,2.3$\msol\citep{kippenhahn2012a} form cores consisting of up to $0.48$\msol of degenerate He. However, we expect the mechanism to work only in a very restricted parameter range.
The formation and survival of a cold, overdense, structure (part of it originating from the core of the He WD) during the circularization of the accreting material seems to be key in the He ignition mechanism.
The survival of  this overdense region is expected to depend sensitively on the masses of the components:
Increasing the CO WD mass leads to higher central densities which favors the CO ignition, but the increased tidal forces  imply that the disruption of the He WD will take place further away from the CO WD and that the disrupted material will be more thoroughly homogenized by the time it is accreted and therefore He ignition may be avoided in the first place.
Decreasing the mass of CO will lower the densities even further and, although the survival chances of the overdense He region could increase, the conditions for the CO detonation will be more difficult to meet. Even if the core is able to detonate, less material will be burned and less $^{56}$Ni will be produced making the event even fainter. Furthermore, as a lower-mass CO WD is less compact, the accreted He at its surface is more unlikely to trigger a detonation. 

In our setup, the He WD had a mass of $0.4$\msun. In principle, a more massive He shell will increase chances of triggering both the He and the CO detonations, but as a core of a RG star, its mass cannot exceed $0.48$\msun \citep{kippenhahn2012a}. In contrast, if the mass of the He WD is reduced, its compactness and central density are reduced and therefore tidal disruption will take place at larger distances and the material will have lower densities, diminishing its chances of forming the overdense structure required for He ignition. 

If the conditions for triggering a detonation are met, the large amount of He mass accumulated on top of the CO WD helps to quickly spread the He detonation around the CO WD. Therefore the convergence of the shockwave is more symmetric than in other scenarios \citep{fink2010a} and therefore the convergence happens more closely to the center of the CO WD. This is important because the density and temperature of this low-mass object are low, and a more central convergence helps to ignite the second detonation.

The progenitor of our considered system has to evolve through a CE phase. Whether this CE interaction leads to a successful ejection of the RG envelope is a topic of ongoing research. Successful envelope ejection was observed in CE simulations of systems such as that considered here only under the assumption of local thermalization of recombination energy, which is challenged by the possibility of energy transport due to radiation or convection \citep[see][ and references therein]{roepke2023a}. If the envelope is successfully ejected, the resulting binary system will consist of a He WD and a CO WD in a tight orbit. The merger and subsequent explosion would then take place after sufficient orbital shrinkage due to the release of gravitational waves. This gives rise to the scenario we have discussed. However, if the envelope is not successfully ejected, the CE event results in the merger between the core of the RG and the companion WD, sometimes called a ``failed CE event'' or a ``CE merger'', see, e.g. \citet{roepke2023a}. 
To test whether our proposed explosion mechanism still triggers an explosion when the secondary star is not a He WD but a RG core, we have carried out a second simulation with the same parameters but a He WD with the initially uniform temperature increased to $2\E{7}\,\textrm{K}$.
This structure is somewhat less compact than the cold He WD in our original simulation. With this altered setup, we still observe the same explosion mechanism, releasing roughly the same nuclear energy and producing equivalent nucleosynthetic yields.
This scenario is reminiscent of the core-degenerate scenario \citep{kashi2011a,ilkov2012a} for SNe~Ia, where the merger between the core of an asymptotic giant branch (AGB) star and a CO WD after a failed CE ejection results in a super-Chandrasekhar mass WD that explodes as a SNe~Ia inside the CE. Our setup differs from the original core-degenerate scenario because our explosion results from a double-detonation of a sub-Chandrasekhar mass CO WD due to the merger with a core of a RG star. With the chosen typical mass of a CO WD, this results in an event much fainter than a SN~Ia.

This simulation shows that it is possible to produce thermonuclear transients in a realistic scenario that synthesise even less $^{56}$Ni and substantially more $^{40}$Ca than what is usually obtained in simulations of the double-detonation scenario aiming at modeling SNe~Ia \citep{boos2021a}.
We reach good agreement with several lightcurves of observed Ca-rich transients, both in shape and peak magnitude. The spectra produced by our model, however, do not reproduce the feature at around $5900$\,\AA\, which is a major discrepancy with the observed spectra.

\citet{zenati2023a} reproduce this feature rather well, while it is entirely missing from our spectra.
The authors argue that their simulation does not have enough remaining He to produce strong He features, suggesting that this feature results from sodium. We find, however, that even increasing the sodium content by a factor of $100$ the spectral appearance remains virtually unchanged.
This suggests that the feature is a He-I line, which \textsc{ARTIS} cannot model in the utilized setup, and the majority of the sodium has been ionized to higher states.
It should be pointed out that although \citet{zenati2023a} reproduce this feature, their simulation differs from ours. The key difference is that in their model, the core of the HeCO WD companion is assumed to survive the event.
This has significant implications on the ejecta structure and composition, in particular regarding the distribution of He in the ejecta and the subsequent potential imprint of He on the spectra.
It remains to be seen if the feature around $5900$\,\AA\ is indeed a He-I line or a Na-I line instead. A strong He-I feature, e.g., would suggest that substantial amounts of He are required in the ejecta, making the disruption of the He-carrying companion the more likely scenario, and vice versa.
Our results so far indicate that this is a He-I line, which is simply beyond the capabilities of our utilized \textsc{ARTIS} setup.

In the nebular phase of Ca-rich transients typically show a high ratio of Ca/O. Our model predicts a substantial amounts of oxygen in the ejecta resulting from unburnt material and from C-burning at low densities in the outer regions of the WD. Whether this oxygen produces a strong feature in the nebular spectra leading to a conflict with observed nebular spectra of Ca-rich transients has to be tested in detailed radiative transfer simulations that are beyond the scope of this paper.
Ca-rich transients are known to also occur far from the centers of their host galaxies\citep{shen2019a}. Whether the progenitor system for our mergers are also located far from the center of galaxies must be tested with population synthesis studies.
Investigating these questions and the nature of the He-I/ Na-I region remain as open tasks in fully assigning our model as a Ca-rich transient progenitor and will be the focus of future work.

In contrast, if the detonation were to take place inside a CE, the spectra and lightcurve would significantly change due to the interaction of the ejecta with the envelope material, showing hydrogen features in the spectra and likely resembling an interacting SN. This interaction has to be modeled in the framework of radiation hydrodynamics and will be further discussed in Kozyreva et al.\footnote{in prep}

The transients resulting from these mergers could be faint, but very frequent depending on the sensitivity of the explosion mechanism to the masses of the components. These explosions could be interesting targets for the Zwicky Transient Facility \citep[ZTF][]{bellm2019a,graham2019a} or the future Vera Rubin Observatory \citep{ivezic2019a}. Detecting these explosions from both, bare core mergers ,and inside an envelope, might also give information about the CE mechanism.

\section*{ACKNOWLEDGMENTS}
J.M-F. and A.H. are fellows of the International Max Planck Research School for Astronomy and Cosmic Physics at the University of Heidelberg (IMPRS-HD) and acknowledge financial support from IMPRS-HD.

This work acknowledges support by the European Research Council (ERC) under the European Union’s Horizon 2020 research and innovation programme under grant agreement No.\ 759253 and 945806, the Klaus Tschira Foundation, and the High Performance and Cloud Computing Group at the Zentrum f{\"u}r Datenverarbeitung of the University of T{\"u}bingen, the state of Baden-W{\"u}rttemberg through bwHPC and the German Research Foundation (DFG) through grant no INST 37/935-1 FUGG.
The authors gratefully acknowledge the Gauss Centre for Supercomputing e.V. (www.gauss-centre.eu) for funding this project by providing computing time through the John von Neumann Institute for Computing (NIC) on the GCS Supercomputer JUWELS at Jülich Supercomputing Centre (JSC).

\bibliographystyle{aa}
\bibliography{astrofritz.bib}
\end{document}